# From Quantum to Classical Physics: The Role of Distinguishability

R. E. Kastner[*]
14 August 2017

ABSTRACT. The transition from quantum to classical statistics is studied in light of Huggett's finding that the empirical data do not support the usual claim that the distinction between classical and quantum objects consists in the capacity of classical objects to carry permutable labels as opposed to quantum objects. Since permutation of the labels of classical objects counts as a distinct configuration, this feature is usually taken as signifying that classical objects are not identical while quantum objects are. Huggett's finding threatens that characterization of the distinction between classical and quantum objects. The various statistical distributions are examined, and it is found that other distinctions, corresponding to separability and distinguishability, emerge in the classical limit. The role of the chemical potential (the rate of change of the Helmholtz free energy with particle number) is found to be of crucial significance in characterizing this emergence of classicality from the quantum distributions.

This paper consists of three main sections. In Section 1, I review Nick Huggett's finding regarding the non-relevance of permutable labeling for the classical/quantum divide (Huggett 1999). In Section 2, I review the derivations of the classical and quantum statistics and argue that a form of *separability* is a key feature of the quantum-to-classical transition. In Section 3, I consider the question: What allows separability to serve as a form of distinguishability in the classical limit? First, let us review some basic considerations regarding issues of individuality and distinguishability.

Steven French (2015) has noted that the concept of *individuality* is primarily a metaphysical isssue, while that of *distinguishability* is primarily an epistemological issue. Nevertheless, distinguishability does have bearing on ontological questions such as:
    What is an individual?
    Are there any true individuals?
    Does Leibniz' Principle of Identity of Indiscernibles apply to Nature?

However, I will not enter here into the metaphysical debate concerning questions such as "what is an individual?" and "are quantum systems individuals?" Rather, I will focus on the issue of *distinguishability* regarding the quantum-classical divide, and attempt to identify some ontological features that may underlie the form of distinguishability obtaining in that context.

---

[*] Foundations of Physics Group, UMCP; rkastner@umd.edu

[1] This concept can be identified with the term 'transcendent individuality' (TI) as discussed in French and Readhead (1988).

1. Huggett's finding on Haecceitism and Classical Objects

Davis Lewis (1986) introduced the term *haecceitism*, which denotes a form of strong individuality: an individual's identity is taken as a primitive "this-ness" which transcends all its qualitative features.[1] (Anti-haecceitism consists in saying that an individual's identity is constituted by its qualitative features and nothing more.) Although the precise definition of haecceitism is still a matter under discussion, for our purposes we can think of it as the capacity of an entity to carry a label or 'name,' where that label is not contingent on any of its qualitative features. Thus, what makes a person named Fred "Fred the individual" is his primitive this-ness, not the color of his eyes or hair or his height, weight, etc.

Now let us consider this notion as applied to some typically classical objects, such as a pair of coins that are assumed to be completely identical and can be either 'heads' or 'tails.' Give them name-labels, say "Fred" and "Joe"-- their assumed haecceitism is represented by their name-labels. In this context, haecceitism implies that if we consider the case in which Fred and Joe are in different states (one of them being 'heads' and the other 'tails'), then interchanging Fred and Joe constitutes two different possible configurations. Including the cases in which Fred and Joe are both 'heads' or both 'tails,' we have four possible states of the coins:

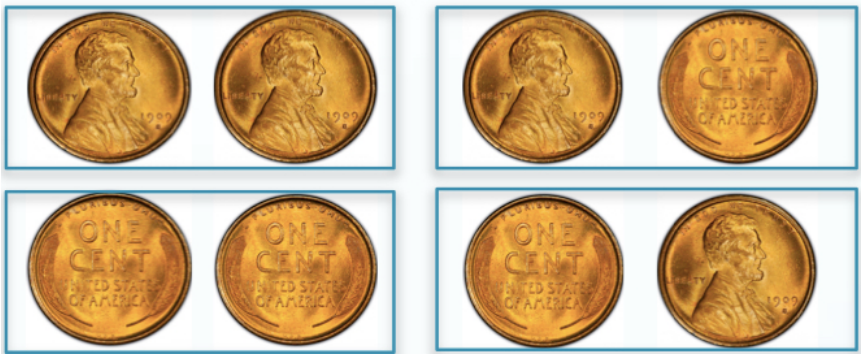

In contrast, for a pair of hypothetical 'boson coins,' the usual story is that there is no such thing as 'Fred' or 'Joe'-- no nameable identity that transcends the qualitative properties of the quantum coins. So the possible configurations are just three in number:[2]

---

[1] This concept can be identified with the term 'transcendent individuality' (TI) as discussed in French and Readhead (1988).

[2] It should be noted that French and Redhead (1988) dissent from this usual identification of individuality with the capacity to carry a label such that permutation of the labels establishes a different state of the total system. They argue that a form of individuality can still be retained for quantum systems if one argues that certain states are not accessible to the total system. For purposes of this discussion, we work with Huggett's formulation, but note that his interpretation of the metaphysical bearing of the labels is not obligatory.

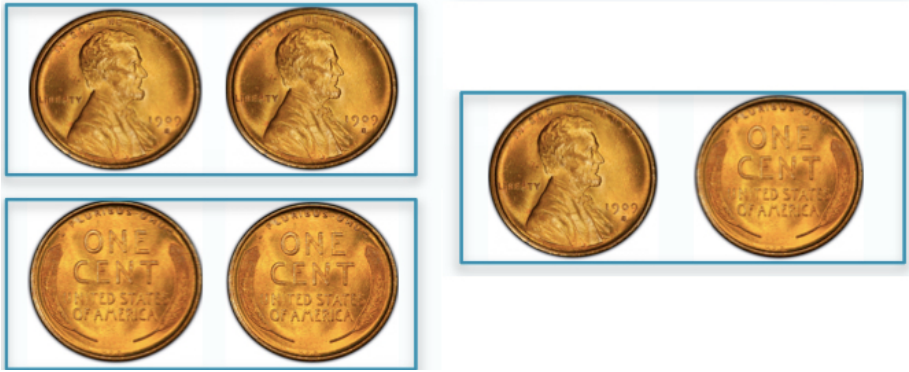

Huggett (1999) notes that there are two ways of describing the state space of a composite system such as the set of two coins. We can either use a phase space (Γ-space) description, which specifies *which* component system is in which state, or we can use a distribution space (Z-space) description, which just specifies how many systems are in each state. The Γ-space description assumes that each component can be meaningfully labeled and/or distinguished from the others, so it supports H-ism in that respect. In contrast, since Z-space specifies only the occupancy number of each state, without identifying any particular system with any particular state, it doesn't support haecceitism in the same way. Since it's typically supposed that the key distinction between classical and quantum objects is the ability of the former to carry a label, one would think that the two kinds of descriptions --phase space and distribution space --would lead to different kinds of statistics; i.e., classical and quantum statistics, respectively.

However, Huggett shows that if we assume that classical objects are impenetrable--i.e. that no more than one such object can never occupy a given spacetime point--then it turns out that the Γ- and Z-space descriptions give the same empirical predictions. Thus, we cannot use any experimental data to decide between them. This means that there is no *empirical* support for the idea that classical and quantum objects differ fundamentally in their metaphysical nature as individuals.

The basic argument goes like this: in terms of the coin analogy, we have to pretend that there are no other qualitative differences between the coins and forbid the two coins 'Fred' and 'Joe' (they can keep their labels) from occupying the same state. Of course, real coins would not fulfill this criterion. For the more realistic case of classical gas molecules, the operative condition is that no two molecules can ever occupy the same individual phase space state, since they can never be at the same spacetime point.

In the case of the idealized coins, if we forbid them from occupying the same state, there are now only two available composite Γ-states for Fred and Joe--the ones in which they are in different 'heads' or 'tails' states:

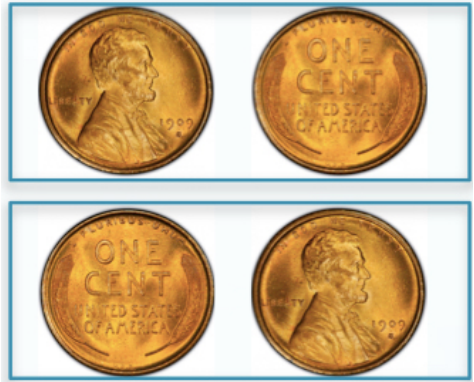

And since both of these correspond to the distribution "one coin in each state", the frequency of this distribution is 2/2 =1. Meanwhile, the frequency of this distribution in terms of the Z-space representation, which ignores the phase space configurations, is just 1/1 =1. We see that, for the idealized classical coins, the frequency of occurrence of the distribution is exactly the same in either representation, so there is no empirical difference between the two spaces—they both predict the same probabilities. Huggett shows that this holds in general, for an arbitrary number of systems and states (i.e., in which the frequency of a given distribution may differ from unity in contrast to the trivial example above).

Thus, it turns out that there is no empirical support for the Γ-space description over the Z-space description for classical systems, if they are correctly characterized as impenetrable)--and thus (IF one identifies that as a criterion for haecceitism as does Huggett), no empirical support for haecceitism as applying to classical objects.[3] While somewhat bewildering for our intuitions about the difference between classical and quantum objects, we actually *need* this result. Why? Because, in keeping with the correspondence principle, the classical (Maxwell-Boltzmann) distribution must (and does) emerge as a limit from quantum statistics (either Bose-Einstein or Fermi-Dirac). That is, the quantum distributions transition smoothly into the classical distribution. This calls into question the idea that classical objects have any sort of "digital" or on/off distinguishability or individuality feature(s) that differ from quantum objects.

Thus, the challenge facing us is that the transition from the quantum domain to the classical domain seems *continuous*, not discrete and essential (as in a change of intrinsic character or essence). This is another puzzling feature of the micro/macro divide. In the next section, we review the derivations of each kind of distribution, and consider what clues we might find therein to better understand the ontology underlying the transition between the quantum and classical statistics.

2. Classical vs. quantum statistics

Let us begin by simply listing the three major distributions; the classical Maxwell-Boltzmann, Bose-Einstein for bosons, and Fermi-Dirac for fermions, respectively:

---

[3] Based on the dissent of French and Redhead (1988) from the above criterion for transcendental individuality, these authors could of course still argue that both quantum systems and classical possess metaphysical individuality. What is off the table, in view of Huggett's argument, is the idea there is any empirical support for a fundamental difference between quantum and classical systems regarding their status as individuals.

$$\bar{n}_i^{(MB)} = N \frac{e^{-\beta\varepsilon_i}}{\sum_j e^{-\beta\varepsilon_j}}$$

$$\bar{n}_i^{(BE)} = \frac{1}{e^{\beta(\varepsilon_i-\mu)} - 1}$$

$$\bar{n}_i^{(FD)} = \frac{1}{e^{\beta(\varepsilon_i-\mu)} + 1}$$

(1a,b,c)

In the quantum distributions (1b) and (1c), the chemical potential $\mu$ (related to the number of degrees of freedom $N$) necessarily enters for systems with a fixed $N$. This will turn out to be significant, as we shall see.

Now, recall that classical distributions can only be wavelike *or* particle-like. The particle-like classical distribution (applying to systems such as ideal gases) is just the Maxwell-Boltzmann distribution (1a). Meanwhile, the classical *wave* distribution is what was applied to blackbody radiation prior to the advent of quantum theory, resulting in the Rayleigh-Jeans distribution and the 'ultraviolet catastrophe' :

$$P(\varepsilon)_{RJ} = \frac{2\varepsilon^2 kT}{(hc)^2} \quad (2)$$

In view of 'wave-particle duality,' it is well known that the quantum distributions interpolate between these two extremes, as follows. Consider the correct quantum distribution for electromagnetic blackbody radiation (the "Planck Distribution"):

$$P(\varepsilon) = \frac{2\varepsilon^3}{(hc)^2} \frac{1}{e^{\frac{\varepsilon}{kT}} - 1} \quad (3)$$

For energies e small compared to $kT$, this becomes

$$P(\varepsilon \ll kT) = \frac{2\varepsilon^3}{(hc)^2} \frac{1}{(1 + \frac{\varepsilon}{kT} + \ldots) - 1} \approx \frac{2\varepsilon^3}{(hc)^2} \frac{kT}{\varepsilon} = \frac{2\varepsilon^2 kT}{(hc)^2} \quad (4)$$

i.e., it yields the Rayleigh-Jeans law (2).

On the other hand, for $\varepsilon$ large compared to $kT$, the exponential in the denominator swamps the unity term and we get

$$P(\varepsilon >> kT) \propto \frac{1}{e^{\frac{\varepsilon}{kT}}} \tag{5}$$

which is the Maxwell-Boltzmann distribution, reflecting particle-like (or at least discrete) behavior on the part of the radiation. (To better reveal the basic form of the distribution in this limit, we neglect the factors corresponding to the density of states for blackbody radiation.)

Thus, we see that the quantum statistics interpolate between wavelike and particle-like behavior. This is a key aspect of quantum systems as opposed to classical systems; the latter can be unambiguously categorized as *either* waves *or* particles. In contrast, the quantum statistics must cover *both* situations in the same distribution--indicating that they describe entities that are (somehow) *both* wave and particle.

If we look at the assumptions that go into deriving the various distributions, we can get some additional clues as to the key differences between the classical and quantum situation. For example, the classical Maxwell-Boltzmann distribution for an ideal gas (e.g., a system of $N$ molecules) is obtained from a partition function assumed to be the direct product of the $N$ individual molecular partition functions. This cannot be done for the quantum statistics for fixed $N$. We will now consider those issues in detail.

First, let's recall the general procedure for obtaining a distribution describing the mean number of systems $\bar{n}_r$ in a given energy state $\varepsilon_r$. This procedure holds regardless of the type of system considered (whether quantum or classical).

- The number of degrees of freedom having energy $\varepsilon_r$ is denoted $n_r$
- Thus, the possible energy states $E_R$ of the whole gas (having $N$ particles) are:

$$E_R = n_1\varepsilon_1 + n_2\varepsilon_2 + \ldots = \sum_r n_r\varepsilon_r ; \tag{6}$$

$$\text{and } N = \sum_r n_r \tag{7}$$

At this point, it may already be noted that (6) represents a distribution over the possible energy states, and in that sense is the "Z-space" representation. Since this is a general derivation (leading to both classical and quantum statistics), it is clear that the Z-space representation is applicable for both cases, reinforcing Huggett's observation.

Now, for the case in which the total system of $N$ degrees of freedom is taken as capable of exchanging energy with its environment at temperature $T$ (the 'canonical ensemble'), the probability that the total system is in the state $R$ is given by

$$P_R = C \exp(-\beta E_R), \tag{8}$$

where $\beta = 1/kT$.

The constant of proportionality $C$ is $1/Z$, where $Z$ is the total system partition function:

$$Z = \sum_j e^{-\beta E_R} \tag{9}$$

So the probability that the gas is in state $R$ is:

$$P_R = \frac{e^{-\beta E_R}}{Z} \tag{10}$$

From this we can find the average number of degrees of freedom in energy state $\varepsilon_r$:

$$\bar{n}_r = \sum_R n_r P_R = \frac{\sum_R n_r e^{-\beta E_R}}{Z} = -\frac{1}{\beta}\frac{1}{Z}\sum_R \frac{\partial}{\partial \varepsilon_r} e^{-\beta \sum_r n_r \varepsilon_r} = -\frac{1}{\beta}\frac{1}{Z}\frac{\partial Z}{\partial \varepsilon_r} \tag{11}$$

So that in compact form,

$$\bar{n}_r = -\frac{1}{\beta}\frac{\partial \ln Z}{\partial \varepsilon_r} \tag{12}$$

Again, this is a general result for any partition function $Z$.

Now, for a *single* degree of freedom with possible energy states $\varepsilon_i$, the partition function $\zeta$ (i.e., the weighted sum over the possible energy states) is given by:

$$\zeta = \sum_j e^{-\beta \varepsilon_j} \tag{13}$$

So, analogously with (10), the probability that a *single* system is in state $\varepsilon_i$ is

$$P(\varepsilon_i) = \frac{e^{-\beta \varepsilon_i}}{\sum_j e^{-\beta \varepsilon_j}} = \frac{e^{-\beta \varepsilon_i}}{\zeta} \tag{14}$$

We make note of this because for a classical gas of $N$ degrees of freedom, one finds the average number simply by taking $Z(N)$ for the entire gas as the product of the individual partition functions:

$$Z(N) = \zeta^N \qquad (15)$$

So that, using (12), the distribution for becomes:

$$\bar{n}_i = -\frac{1}{\beta}\frac{\partial \ln Z(N)}{\partial \varepsilon_i} = -\frac{1}{\beta}N\frac{\partial \ln \zeta}{\partial \varepsilon_i} = N\frac{e^{-\beta \varepsilon_i}}{\sum_j e^{-\beta \varepsilon_j}} \qquad (16)$$

which is just the Maxwell-Boltzmann distribution (1a).

However, one cannot use the expression (15) for quantum systems that have a constrained number of degrees of freedom $N$--and this is of crucial significance. Instead, one must incorporate the restriction to $N$ by way of the chemical potential $\mu$, which acts as a Lagrange multiplier. This dictates that we are working with the "grand canonical ensemble," which allows $N$ to vary. The corresponding grand canonical partition function $\check{Z}$ is obtained as follows:

$$\check{Z} = \sum_R \exp[-\beta E_R]\exp(\beta \mu N) = \sum_{n_1, n_2, n_3 \ldots} \exp[-\beta(n_1 \varepsilon_1 + n_2 \varepsilon_2 + n_3 \varepsilon_3 + \ldots)]\exp(\beta \mu N) =$$

$$= \sum_{n_1, n_2, n_3 \ldots} \exp[-\beta(n_1 \varepsilon_1 + n_2 \varepsilon_2 + n_3 \varepsilon_3 + \ldots)]\exp[\beta\mu(n_1 + n_2 + n_3 + \ldots)]$$

$$= \sum_{n_1, n_2, n_3 \ldots} e^{-\beta n_1(\varepsilon_1 - \mu)}e^{-\beta n_2(\varepsilon_2 - \mu)}e^{-\beta n_3(\varepsilon_3 - \mu)}\ldots = \left(\sum_{n_1=0}^{\infty} e^{-\beta n_1(\varepsilon_1 - \mu)}\right)\left(\sum_{n_2=0}^{\infty} e^{-\beta n_2(\varepsilon_2 - \mu)}\right)\left(\sum_{n_3=0}^{\infty} e^{-\beta n_3(\varepsilon_3 - \mu)}\right)\ldots \qquad (17)$$

This is just a product of infinite sums of the form $\sum_{n=0}^{\infty} x^n = \frac{1}{1-x}$, $|x|<1$; and given that $\mu < \varepsilon_r$, we therefore have

$$\check{Z} = \left(\frac{1}{1-e^{-\beta(\varepsilon_1 - \mu)}}\right)\left(\frac{1}{1-e^{-\beta(\varepsilon_2 - \mu)}}\right)\left(\frac{1}{1-e^{-\beta(\varepsilon_3 - \mu)}}\right)\ldots \qquad (18)$$

Taking logs of both sides, we get the more useful form:

$$\ln \check{Z} = -\sum_{r=0}^{\infty} \ln\left(1 - e^{-\beta(\varepsilon_r - \mu)}\right) \qquad (19)$$

And then can use (12) to get the distribution for average occupation number $\bar{n}_s$:

$$\bar{n}_s = -\frac{1}{\beta}\frac{\partial \ln \check{Z}}{\partial \varepsilon_s} = -\frac{1}{\beta}\frac{\partial}{\partial \varepsilon_s}\left[-\sum_{r=0}^{\infty}\ln\left(1-e^{-\beta(\varepsilon_r-\mu)}\right)\right] = \frac{e^{-\beta(\varepsilon_s-\mu)}}{1-e^{-\beta(\varepsilon_s-\mu)}} = \frac{1}{e^{\beta(\varepsilon_s-\mu)}-1}, \quad (20)$$

which is the Bose-Einstein distribution.

The first thing to notice here (besides the fact that we could not use (15) to obtain this quantum distribution) is that the total number of degrees of freedom, $N$, seems to have 'disappeared.' It got 'dissolved' into infinite sums over all the possible values of the $n_s(\varepsilon_s)$. Thus, ironically, *N has to become a variable in order to be able to 'fix' N for a gas of quantum systems.* We recover $N$ as the sum over the average occupation numbers $n_s$:

$$N \equiv \bar{N} = \sum_s \bar{n}_s \quad (21)$$

The situation is similar for fermions, except that they obey the Pauli Exclusion principle which limits state occupancy to zero or one. Without going through the derivation here, we note given the above restriction on occupancy, the inability to express the partition function $Z(N)$ as a direct product of $N$ individual degrees of freedom yields for the mean occupancy number:

$$\bar{n}_s^{FD} = \frac{1}{e^{\beta(\varepsilon_s-\mu)}+1} \quad (22)$$

which is the Fermi-Dirac distribution.

Thus, for both bosons and fermions, the chemical potential $\mu$ is involved in a crucial, non-separable way. Its relation to $N$ is fixed by (21), i.e.:

$$N = \bar{N} = \sum_s \frac{1}{e^{\beta(\varepsilon_s-\mu)}\pm 1} \quad (23)$$

Does the chemical potential play any role in the classical case? Yes, but only trivially, as a normalizing factor. In the 'dilute' (low-occupancy) limit yielding the classical case, the exponential factor involving $\mu$ approaches the particle number $N$ divided by the single-particle partition function, i.e.,

$$e^{\beta\mu} \rightarrow \frac{N}{\zeta} = \frac{N}{\sum_j e^{-\beta\varepsilon_j}} \quad (24)$$

So that the Maxwell Boltzmann distribution can be expressed in terms of $\mu$ as:

$$\bar{n}_i^{(MB)} = \frac{N}{\zeta} e^{-\beta\varepsilon_i} = e^{\beta\mu} e^{-\beta\varepsilon_i} = \frac{1}{e^{\beta(\varepsilon_i - \mu)}} \tag{25}$$

In this form, it is easy to see that the classical case emerges from the quantum distributions when $\varepsilon_i \gg \mu$ for all i.

In summary, we make the following observations based on the derivations of the respective distributions. For a classical system comprising *N* degrees of freedom, we can simply assume that *N* is fixed, and use the 'canonical ensemble' to obtain the distribution. For that purpose, we can express the total canonical partition function *Z(N)* as simply the product of the individual partition functions $\zeta_i$ for the component degrees of freedom.

But for a quantum system with fixed *N* (and nonvanishing mass), we cannot use the canonical ensemble; we *must* use the 'grand canonical ensemble' (i.e. partition function Ž)– representing a system in contact with both an energy and particle reservoir. That is, we must in-principle allows *N* to vary. The physical content of this procedure is as follows: the chemical potential $\mu$ is a Lagrange multiplier, representing a constraint force that is present in the quantum case, even if there is no contact with an external particle reservoir. Thus the natural physical interpretation is that *the quantum degrees of freedom are imposing a constraint force on one another that is not present in the classical case*.

Based on the above, what can we conclude about the classical/quantum divide? We cannot treat a collection of *N* quantum objects as elements of separable probability spaces, since in that case we do not obtain the QM statistics. Non-separability of the spaces confirms that we are dealing with quantum coherence, with all its attendant features such as entanglement and the requirement for symmetrization rendering labels superfluous (the latter usually and reasonably understood as reflecting indistinguishability). Moreover, the mutual constraint of quantum systems expressed by the chemical potential $\mu$ (even at T=0) reflects a peculiarly quantum sort of physical correlation or interaction not present in the classical case (probably expressing the so-called 'exchange forces' associated with symmetrization).[4]

Thus, our finding is that there is no empirical support for any 'digital' on/off form of metaphysical individuality at the classical/quantum border. Since the classical statistics are straightforwardly obtainable as a limit from the quantum statistics, and representable in terms of *Z*-space, we can confirm Huggett's result that the capacity of classical systems to carry labels that simply permute to form new system states (i.e., new Γ-space configurations) is not reflected in the statistics. Yet clearly, the classical limit brings with it some sort of new capacity for permutable labeling of the component systems, in that the collective partition function can be obtained from individual partition functions $\zeta_i$ in-principle capable of carrying the permutable label *i*. This indicates that in the classical limit, the component systems acquire a form of distinguishibility. In the next section, we investigate the nature of this emergence, in the classical limit, of the capacity to carry a label.

---

[4] Of course, this is a misnomer; there is no real 'force' operating here in the usual physical sense.

3. Whence quasi-classical distinguishability in the dilute limit?

The dilute limit, yielding classicality, is known to be obtained in the 'small wavelength limit,' through the use of the so-called 'thermal wavelength' $\lambda_{th}$. The usual ways of obtaining the $\lambda_{th}$ condition can be criticized for conflating classical and quantum quantities. For example, one typical method for deriving $\lambda_{th}$ is by treating it as a kind of quantum-mechanical position uncertainty $\Delta x$ corresponding to the root-mean-square uncertainty $\Delta p^{RMS}$ of the momentum of the component degrees of freedom (at a given temperature T). That is, one starts with the expression

$$\Delta p^{RMS} = \sqrt{\langle p^2 \rangle - \langle p \rangle^2} \qquad (26)$$

The average momentum <p> appearing in (26) is assumed to be zero because of 'random motion'-- thus, it is an average over *N* independent degrees of freedom, not an expectation value for any quantum state. $\Delta p^{RMS}$ is then taken as equal to the square root of the average squared momentum <$p^2$>, which is obtained from the equipartition theorem:

$$\langle p^2 \rangle = 3mkT \qquad (27)$$

So from (26) and (27), the quantity $\Delta p^{RMS}$ is taken to be

$$\Delta p^{RMS} = \sqrt{3mkT} \qquad (28)$$

and this (despite the fact that it is not a real momentum uncertainty but rather a root-mean-square error) is plugged into the uncertainty relation to obtain a corresponding 'thermal wavelength' $\lambda_{th}$ :

$$\lambda_{th} = \frac{h}{\sqrt{3mkT}} \qquad (29)$$

Clearly, in this context, $\lambda_{th}$ is assumed to be a kind of position uncertainty. It's then demanded that this be much smaller than the average interparticle spacing *d*, where

$$d = \left(\frac{V}{N}\right)^{1/3}, \qquad (30)$$

the idea being that this condition makes the gas "dilute" (i.e., no particles ever occupying same position *x*; and many positions unoccupied--see Figure 1.). So the 'thermal wavelength' condition for classicality becomes:

$$\frac{h}{\sqrt{3mkT}} << \left(\frac{V}{N}\right)^{1/3} \tag{31}$$

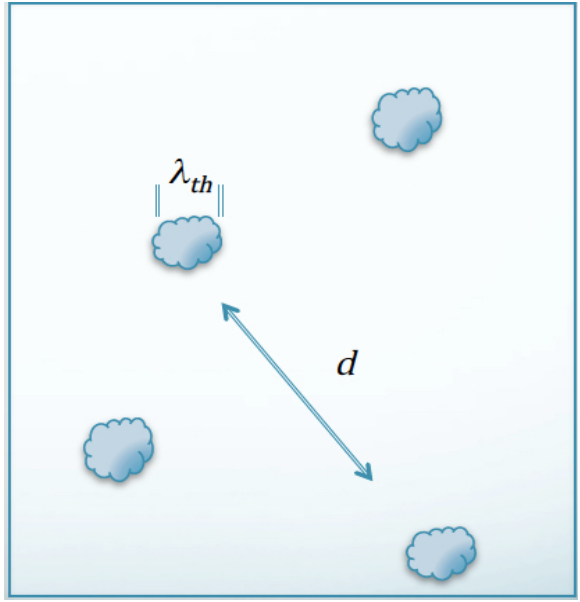

Figure 1. One way of picturing the 'thermal wavelength' condition for the classical limit of a quantum distribution.

However, as alluded to above, the preceding derivation applies a *classical*, not quantum, uncertainty to momentum in two distinct ways:

- averaging over *N* systems assumed to be in 'random motion' to get a vanishing value for *<p>*
- use of the classical Equipartition Theorem to get a value for *<p²>*

The resulting momentum uncertainty is only an epistemic average over many different phase space points, not an intrinsic quantum uncertainty arising from a single state having an intrinsic 'coarse-graining' represented by a finite-sized element of phase space. It is arguably therefore not the correct quantity for applicability of the uncertainty relation between position and momentum. Moreover, the derivation of the Equipartition Theorem, which is used to obtain that root-mean-square 'momentum uncertainty,' *presupposes classical* Maxwell-Boltzmann statistics!

In addition, the derivation conflates a wavelength with a position uncertainty, which is problematic: if a system has a well-defined wavelength $\lambda$, then it has infinite position uncertainty. Thus, $\lambda_{th}$ is really being interpreted as the spread of a 'wave packet,' despite the fact that the states available to the systems making up the gas are not necessarily described by wave packets (i.e., in the quantum limit, they are plane waves).

Of course, it is well-known that classical behavior emerges in the small-wavelength limit, so rather than try to pretend that a wavelength is a position uncertainty, one can simply work with the DeBroglie wavelength of the average momentum (still using the Equipartition Theorem) to obtain the same condition (31). But in this case, one cannot explain the classical behavior in this limit by saying that the gas is 'dilute,' as pictured in Figure 1, since it retains a nonlocal character arising from the presumed exact wavelengths of its degrees of freedom. So the question is: why does condition (31) seem to work so well as a criterion for the classical limit?

It turns out that if we re-express (31) in terms of thermal energy $kT$, we find the condition (neglecting numerical constants of order unity):

$$kT \gg \frac{h^2}{m}\left(\frac{N}{V}\right)^{2/3} \qquad (32)$$

But the quantity on the right-hand side is then recognized as the Fermi Energy, which is the chemical potential $\mu$ for fermions at T=0:

$$E_F = \frac{h^2}{m}\left(\frac{N}{V}\right)^{2/3} = \mu(T=0) \qquad (33)$$

And in fact, this condition $kT \gg E_F$ is also the well-known condition for the classical limit of the Fermi-Dirac distribution. So what happens in this limit that could justify a classical description? Once again, the chemical potential $\mu(T)$ has much to teach us.

Specifically, $\mu$ is crucially related to the Helmholtz free energy F, defined as

$$F = U - TS \qquad (34)$$

Specifically, $\mu$ is the change in F when adding a degree of freedom (at a given T and V), *i.e.*:

$$\mu = \left(\frac{\Delta F}{\Delta N}\right)_{T,V} \qquad (35)$$

Now, $\mu$ is of large magnitude and negative in the classical limit of large T, small N/V, and small $\lambda_{th}$, as can be seen from the well-known relation[5]:

$$\mu \xrightarrow[clas]{} -kT \ln\left(\frac{V}{N\lambda_{th}^3}\right) \qquad (36)$$

---

[5] See, e.g., Kelly (2002) for details.

If we can understand the physical significance of the negativity (and large magnitude) of $\mu$ in the classical limit as expressed by (36), we may hope to gain insight into the ontology of the quantum/classical transition. A large negative value of $\mu = \Delta F/\Delta N$ means that $F$ decreases significantly when a degree of freedom is added to the system. Since $F = U - TS$, increasing the entropy S is the only way to decrease $F$. This indicates that the addition of a degree of freedom in the classical limit increases the entropy far more than it increases the internal energy $U$. The higher the temperature T, the larger this decrease, and the more negative $\mu$ becomes. *Thus, a large negative $\mu$ corresponds to a large entropy increase of the whole system whenever a degree of freedom is added, accompanied by a negligible increase in internal energy*. The same basic situation applies to bosons, although in that case $\mu$ can never be positive, and with increasing T, it becomes more and more negative while more of the higher energy states $\varepsilon_s$ become populated (and the system becomes more dilute in terms of state occupancy).

We can therefore summarize as follows, guided by the clues provided by the behavior of the chemical potential. In the classical (dilute) limit, adding a degree of freedom results in a statistically independent increase in the overall state space, increasing the entropy with comparatively small increase in $U$. In contrast, in the quantum domain, we have two cases: (i) for fermions, adding a degree of freedom increases $U$ more than it increases $TS$; (ii) for bosons, the entropy term ($-TS$) is always larger in magnitude than $U$, but the increase in $U$ is non-negligible compared to the increase in the magnitude of $TS$. The physical origin of the relatively small increase in $TS$ when adding a degree of freedom in the quantum limit is the following: the new degree of freedom has to find an energy level contingent on the pre-existing energy level structure, which reduces the availability of states that would have been available in the classical case. Thus the entropy increase (which is a measure of the increase in the number of available states) is much smaller than what obtains in the classical case. Once again, the quantum degrees of freedom "know about each other" and evidently have some form of interaction (quantified by the chemical potential), even at T=0 when there are no thermal interactions at all. They are not independent and separable.

4. Conclusions.

By examining the derivations of the quantum and classical distributions, we have found that separability of the individual probability spaces fails in the quantum domain. In contrast, in the classical limit, the probability spaces of the component degrees of freedom are fully separable. In addition, by examining the role of the chemical potential $\mu$, we find a clear manifestation of the highly non-classical constraints that quantum degrees of freedom impose on one another via the 'exchange forces' corresponding to the need for symmetrization. We also confirm, via the behavior of $\mu$, that in the classical limit, adding a degree of freedom gives rise to new energy states for the whole system of $N$ degrees of freedom, independently of the state occupancies of the pre-existing $N-1$ degrees of freedom—increasing entropy with minimal increase in internal energy.

Thus, classical systems (those obeying Maxwell-Boltzmann statistics) have a form of separability and independence not applying to quantum systems. This separability amounts to distinguishability, since one could in-principle apply labels to the $N$ individual state-spaces making up the collective state space (as in a $\Gamma$-space representation).

However, it is notable that impenetrability does not come into the picture in any fundamental way, because we need only consider energy states (not position) in order to obtain Maxwell-Boltzmann statistics. Nevertheless, what about our intuition (also reflected in the usual derivation of the 'thermal wavelength' criterion as illustrated in Figure 1) that classical objects do not overlap in spacetime, and are fundamentally independent from one another? Einstein addressed this classical notion of separability in terms of his 'being thus' concept:

> An essential aspect of this arrangement of things in physics is that they lay claim, at a certain time, to an existence independent of one another, provided that these objects 'are situated in different parts of space.' Unless one makes this kind of assumption about the independence of the existence (the 'being–thus') of objects which are far apart from one another in space—which stems in the first place from everyday thinking—physical thinking in the familiar sense would not be possible. It is also hard to see any way of formulating and testing the laws of physics unless one makes a clear distinction of this kind. (Einstein, 1948)

Of course, when he made this statement, Einstein was resisting the quantum nonlocality and/or nonseparability that was evident in the context of the famous EPR experiment (Einstein, Podolsky, Rosen 1935). It has since become clear that it is indeed possible, and necessary, to formulate and test the laws of physics without relying on this sort of classical picture at all levels.

We can trace the emergence of Einstein's 'being thus' in the classical limit by noting that the latter obtains for high thermal energies $kT$ (eqn. 32). What can be deduced from that depends on one's interpretation of the quantum formalism. In a unitary-only account, high thermal energies enable decoherence arguments to proceed (Joos and Zeh, 1985), although that account has been criticized on the basis of entanglement relativity and circularity (e.g., Dugić and Jeknić-Dugić (2012), Fields (2010), Kastner 2014, 2016).

Another interpretive approach is to take the projection postulate of von Neumann as a real physical process (i.e. a 'collapse' interpretation). One such approach is actually a different theory from quantum mechanics: the Ghirardi-Rimini-Weber (1985) 'GRW' mechanism requires an *ad hoc* modification to the Schrödinger evolution. In contrast, a collapse interpretation that does not change the basic quantum theory is the (Relativistic) Transactional Interpretation (RTI), which takes the advanced states as playing a physical role in measurement by breaking the linearity of the evolution and giving rise to the von Neumann 'measurement transition' (Kastner 2016a; 2012, Chapter 3).[6] High thermal energies $kT$ give rise to frequent inelastic scatterings among the degrees of freedom of the gas and thermal photons. According to RTI, inelastic scatterings correspond to collapses, which serve to localize the component degrees of freedom—giving them effective separate and distinct spacetime trajectories, conferring independence and thus restoring Einstein's notion of 'being thus'.[7]

---

[6] The original TI as proposed in Cramer (1986) was limited to the non-relativistic domain, and took emitters and absorbers as primitive. The extension of TI to the relativistic domain (RTI) by the present author has allowed a quantitative definition of emitters and absorbers from underlying principles (Kastner 2012, Chapter 6; Kastner 2016a), and full refutation of the consistency challenge raised by Maudlin (1996) The refutation is presented in Kastner 2016b.

[7] Another advantage of this approach is to provide a physical grounding for the Second Law of Thermodynamics at the micro-level (Kastner 2017).

It should also be noted that under the RTI model with a real non-unitary transition defining 'measurement,' interference truly does disappear upon measurement, in contrast to the usual assumption of unitary-only evolution. This resolves the issue alluded to (for example) in French and Redhead (1988), who say:

> But, of course, ontologically speaking, 'interference' is never strictly absent. That, after all, is what constitutes the 'problem of measurement' in QM, so the involvement of every electron with the state of every other electron in the universe, although negligible for practical purposes, remains an ontological commitment of QM, under the interpretation where the particles are treated as individuals. (French and Redhead 1988, 245)

In contrast, under RTI, the above form of global interference does vanish upon the non-unitary transition in which a transaction is actualized. Since transactions are very frequent in the conditions defining the classical limit, this can be seen as directly supporting the independence, or 'being thus,' of systems in the classical limit.

Whichever interpretation one adopts, in the domain of high thermal energies $kT$, one gets at least effective determinacy of position over time, and thus a unique spacetime trajectory for each degrees of freedom. Such a trajectory confers the capacity for a unique label, and therefore supports distinguishability of the degree of freedom to which it corresponds. This does not amount to a haecceitistic label, since it is conferred based on qualitative features of the degrees of freedom (i.e. their trajectories). Nevertheless, as noted by French and Redhead, one may still regard all systems (classical and quantum) as haecceitistic under a suitable interpretation of individuality. This paper takes no position on that metaphysical issue.

Acknowledgments. The author is grateful for valuable correspondence from Jeffrey Bub and Steven French.